\documentclass[preprint,3p,12pt]{elsarticle}

\usepackage{amsmath,amsfonts,amsthm,amssymb, bm}
\usepackage{epsfig}
\usepackage{verbatim,epic, eepic}
\input amssym.def
\input amssym.tex
\usepackage{yfonts}[1998/10/03]

\newfont{\go}{ygoth.tfm scaled 1200}

\newcommand{\abs}[1]{\left| #1 \right|}

\newcommand{\pd}[2]{\frac{\partial#1}{\partial#2}}
\newcommand{\pdd}[2]{\frac{\partial^2 #1}{\partial #2 ^2}}

\numberwithin{equation}{section}
\journal{Annals of Physics}

\begin{document}

\begin{frontmatter}
\title{Traffic Noise and the Hyperbolic Plane}
\tnotetext[t12]{DAMTP pre-print no. DAMTP-2009-75}
\author[DAMTP]{G.~W.~Gibbons}
\ead{G.W.Gibbons@damtp.cam.ac.uk}

\author[DAMTP,Qns]{C.~M.~Warnick}
\ead{C.M.Warnick@damtp.cam.ac.uk}

\address[DAMTP]{DAMTP, CMS, Wilberforce Road, Cambridge, CB3 0WA, UK}
\address[Qns]{Queens' College, Cambridge, CB3 9ET, UK}
\begin{abstract}

We consider the problem of sound propagation in a wind. We note that
the rays, as in the absence of a wind, are given by Fermat's principle
and show how to map them to the trajectories of a charged particle
moving in a magnetic field on a curved space. For the specific case of
sound propagating in a stratified atmosphere with a small wind speed
we show that the corresponding particle moves in a constant magnetic
field on the hyperbolic plane. In this way we give a simple
`straightedge and compass' method to estimate the intensity of sound
upwind and downwind. We construct Mach envelopes for moving
sources. Finally, we relate the problem to that of finding null
geodesics in a squashed anti-de Sitter spacetime and discuss the $SO(3,1)\times
\mathbb{R}$ symmetry of the problem from this point of view.

\end{abstract}


\end{frontmatter}

\section{Introduction}

In the New Scientist\footnote{issue 2704} of 15$^{th}$ April, 2009, a correspondent posed the following question regarding traffic noise

\begin{quotation}
I live a kilometre north of a busy motorway. When the wind is coming from the south the noise of the motorway is noticeably greater than when the wind is coming from the north.

Assuming a wind speed of a mere 30 kilometres per hour, how can the wind direction affect the level of traffic noise I hear when the speed of sound is more than 1235 kilometres per hour?
\end{quotation}

This apparent paradox, and its resolution, have been known since at
least the time of Stokes \cite{Stokes}. The explanation given by
Stokes is that this effect is produced by \emph{wind shear}, the variability in the wind speed as a function of height. This gives rise to refraction, causing sound rays to bend away from the ground in the upwind direction and towards the ground in the downwind direction.

As a qualitative explanation of the process, this is perfectly
satisfactory, see for example the discussion of Reynolds
\cite{Reynolds}. A quantitative discussion is given by Rayleigh \cite{Rayleigh}
however he assumes that the rays are normal to the wavefronts, which
is not true in the presence of a wind. Rayleigh thus arrives at an
incorrect equation for sound rays in a wind, as pointed out by later
authors \cite{Milne, Kornhauser}.

Modern approaches to this problem usually involve the
use of numerical ray-tracing to plot the paths of sound rays
\cite{Jones}. The purpose of this paper is to provide a quantitative,
analytic, discussion of this effect. We consider a stratified
atmosphere whose sound speed and wind speed are allowed to vary with
height. Making use of geometrical ideas discussed in
\cite{Gibbons:2008zi} we show that provided the wind speed is small
compared to the speed of sound, and that the sound speed does not vary
rapidly with height, the rays are well approximated by trajectories of
a charged particle moving in a uniform magnetic field on the
hyperbolic plane. 
 
We use the equivalence between sound rays and charged particles to
investigate two problems. One is that of traffic noise, outlined
above. The other is the problem of the Mach envelope of a moving body
in a stratified atmosphere with a small wind. We finally discuss how
to lift the problem to that of finding null geodesics for a squashed
anti-de Sitter spacetime, which allows us to exhibit the $SO(3,1)\times
\mathbb{R}$ symmetries in a simple fashion.

\section{Fermat's principle for moving media \label{finssec}}

Consider a disturbance $u(x, t)$ obeying the wave equation
\begin{equation}
\left[\pdd{}{t} - h^{ij} \pd{}{x^i} \pd{}{x^j}
\right]u(x,t) = 0,
\end{equation}
where $h^{ij}(x)$ is a positive definite symmetric matrix. A sound
wave propagating locally in the direction $n^i$ will have phase
velocity $h_{ij}n^i n^j$ at $x$. Here $h_{ij}$ is the matrix inverse
of $h^{ij}$. For the
case of an isotropic fluid, we have $h^{ij} = c^2(x)
\delta^{ij}$, with $c$ the local speed of sound, but we will allow for the
possibility that the speed of sound depends on the direction of
propagation.

Suppose we take a disturbance whose wavelength is short compared to
all other length-scales. It is well known that the
energy of the wave travels along \emph{rays} $\alpha = \alpha(\tau)$, which are extremals of
the funtional
\begin{equation}
T[\alpha] = \int \sqrt{h_{ij} \alpha'^i \alpha'^j} d\tau
\end{equation}
Where $\tau$ is \emph{any} parameter along the curve. We note that the integrand
is simply $dt$, so that rays obey \emph{Fermat's principle of least time}, i.e. a sound packet will
travel from $P$ to $Q$ along a path which minimises\footnote{strictly
  \emph{extremises} -- the ray may in fact maximise the time taken} the time taken
among nearby paths, subject to the condition that the packet moves
always at the local speed of sound.

Now let us consider a disturbance in a fluid with local sound speed
tensor $h^{ij}(x)$ which is moving with a local velocity
$W^i(x)$. This will obey a modified wave equation:
\begin{equation}
\left[\left(\pd{}{t}-W^i \pd{}{x^i}\right)^2 - h^{ij} \pd{}{x^i} \pd{}{x^j}
\right]u(x,t) = 0.\label{wem}
\end{equation}
What is perhaps less well known is that the sound rays in this
situation also obey a form of Fermat principle: a sound packet will
travel from $P$ to $Q$ along a path which minimises the time taken
among nearby paths, subject to the condition that it moves at the
local speed of sound, \emph{relative to the flow}.

This means that a sound ray in a moving medium must solve a Zermelo
navigation problem. This problem, proposed by Zermelo is to find the
path between two points which minimises the time travelled, given that
one moves at unit speed with respect to a `wind' vector $W$. Taking
the view that $h_{ij}$  defines a metric, with respect to which sound moves
at unit speed, the principle of least time tells us that a ray indeed
solves a Zermelo problem.

The general Zermelo problem of navigation on a
manifold with metric $h$ and wind $W$ is considered in \cite{BaoRoblesShen}. It is shown that this problem is
isomorphic with finding the Finsler geodesics of a metric in the Randers class
of Finsler metrics. Randers metrics take the form
\begin{eqnarray}
F &:& TM \setminus 0  \to \mathbb{R}^+\nonumber \\
&& (x,y)  \mapsto  \sqrt{a_{ij}(x)y^i y^j}+b_iy^i \label{fins}
\end{eqnarray}
Where $a_{ij}b^ib^j<1$ is required in order that this be a good
Finsler metric. The Randers metric whose geodesics solve the Zermelo problem with
data $(h_{ij}, W^i)$ is given by
\begin{equation}
a_{ij} = \frac{\lambda h_{ij}+W_i W_j}{\lambda^2}, \quad b_i =
-\frac{W_i}{\lambda}, \quad W_i = h_{ij}W^j, \quad \lambda = 1-h_{ij}W^iW^j.\label{ztor}
\end{equation}
The condition $a_{ij}b^ib^j<1$ becomes $h_{ij} W^iW^j<1$. This
transformation is invertible and every Randers metric may be 
thought of as arising from a Zermelo problem. Geodesics of the Randers
metric are (up to parameterization) the paths followed by a particle
of unit charge, moving at unit speed in the magnetic field defined by
$F = db$. For this reason, we will identify Randers one-forms
which differ by an exact form $b \sim b + d\phi$ for most of this paper, the exception
coming in Section \ref{mach} where the parameterization will be relevant.

The fact that the rays of (\ref{wem}) obey Fermat's principle can be
seen in two ways. In \cite{MeyerSchroeter} it is shown that one may start with the
Hamiltonian system following from the dispersion relation of
(\ref{wem}) and show that the integral curves of the Hamiltonian obey
the Euler-Lagrange equations of the time functional
\begin{equation}
T[\alpha] = \int \sqrt{a_{ij} \alpha'^i \alpha'^j} +b_i  \alpha'^i ds
\end{equation}
with $a$ and $b$ defined in terms of $h$ and $W$ by
(\ref{ztor}). Since the Hamiltonian is homogeneous of degree $1$ in
momenta this transformation is not the standard Legendre
transformation. Alternatively, in \cite{White} Fermat's principle is derived by
considering (\ref{wem}) as the d'Alembertian operator of a
$4$-dimensional Lorentzian manifold. Using the fact that the rays are
determined by the principal part of the wave equation (which is
conformally invariant) one may derive Fermat's principle by
considering the null geodesics of this Lorentzian spacetime. In  \cite{Gibbons:2008zi}  the
relation between the Zermelo, Randers and spacetime viewpoints was
explored in greater depth. 

The link between magnetism and wave propagation in a moving background
has also been explored in a different context by Berry et
al. \cite{Berry}. Here the relation is used to construct a water wave
analogue for a quantum mechanical system.

\section{Traffic Noise \label{trafnoise}}

We consider the following problem as a model for traffic noise in the
vicinity of a motorway, with a stratified atmosphere such that the speed of sound varies with height
above the ground and there is in addition a cross wind, also
varying with height. We work in the upper half plane, with coordinates
$x,z$ and take the ground to be $z=0$. We suppose the sound waves
travel at a speed $c(z)$ and there is a horizontal wind $W = w(z)
\partial/\partial x$ which vanishes at $z=0$. Thus, we seek to solve the Zermelo problem with data
\begin{equation}
h = \frac{dx^2+dz^2}{c^2(z)}, \qquad W = w(z) \pd{}{x}.
\end{equation}
The corresponding Randers data are
\begin{equation}
a = \frac{dx^2}{c^2(z)\left(1-\frac{w^2(z)}{c^2(z)}
  \right)^2}+\frac{dz^2}{c^2(z)\left(1-\frac{w^2(z)}{c^2(z)} \right)},
  \qquad b=-\frac{w(z)}{c^2(z)-w^2(z)}dx.
\end{equation}

If we make the assumptions that $w(z)/c(z) \ll 1$ and $c''(z) c(z)/c'(z)^2\ll
1$, then the Gaussian curvature of the metric $a$ is
\begin{equation}
K \approx -(c'(z)^2+2 w'(z)^2).
\end{equation}
We will therefore approximate the metric $a$ in the neighbourhood of
$z=0$ by a metric of constant negative curvature $K =
-(\sigma_c^2+2 \sigma_w^2)$, where we define for convenience
$\sigma_c = c'(0)$, $\sigma_w = w'(0)$. The line $z=0$ may be shown to have geodesic
curvature
\begin{equation}
k_g = \sigma_c
\end{equation}
so in our approximation, the surface of the ground will be a curve of constant geodesic
curvature in the hyperbolic plane. $\sigma_c$ is positive
when the speed of sound increases with height from ground level and
negative if the speed of sound decreases with height. In the model
proposed, the sign of
$\sigma_c$ determines on which side of the curve of constant geodesic
curvature the atmosphere lies. 

To first order in $w/c$, we find
\begin{equation}
F = db = w'(z) \mu_a
\end{equation}
where $\mu_a$ is the area form of the metric $a$. Thus in our
approximation we will consider the sound rays to be the paths of a
particle or unit charge, moving at unit speed on the hyperbolic plane
with respect to a constant magnetic field $\sigma_w$.

We use these observations to map the problem to one in the
hyperbolic disk\footnote{In a later section,
we will deal with the precise mappings which take our problem over to
the hyperbolic disk, but we already have sufficient information to
propose a geometric construction to determine received intensity as a
function of distance from the source.}. This is the region $\abs{\zeta}<1$ of $\mathbb{C}$,
endowed with the Poincar\'e metric
\begin{equation}
ds^2 = \frac{4 \rho^2}{\left(1-\abs{\zeta}^2\right)^2} d\zeta d\bar{\zeta}.
\end{equation}
This is a model of the hyperbolic plane, with constant curvature $K =
-1/\rho^2$. A particle of unit charge moving in a constant magnetic
field $\beta$, passing through the origin, moves along an arc of a
circle of Euclidean radius
\begin{equation}
R_\beta = \frac{1}{2 \beta \rho}.
\end{equation}
Alternatively, we can characterise this arc as a curve of constant
geodesic curvature $\beta$.

\begin{figure}[!h]
\begin{minipage}[t]{0.46\linewidth} 
\centering {\includegraphics[height=2.5in,
width=2.5in]{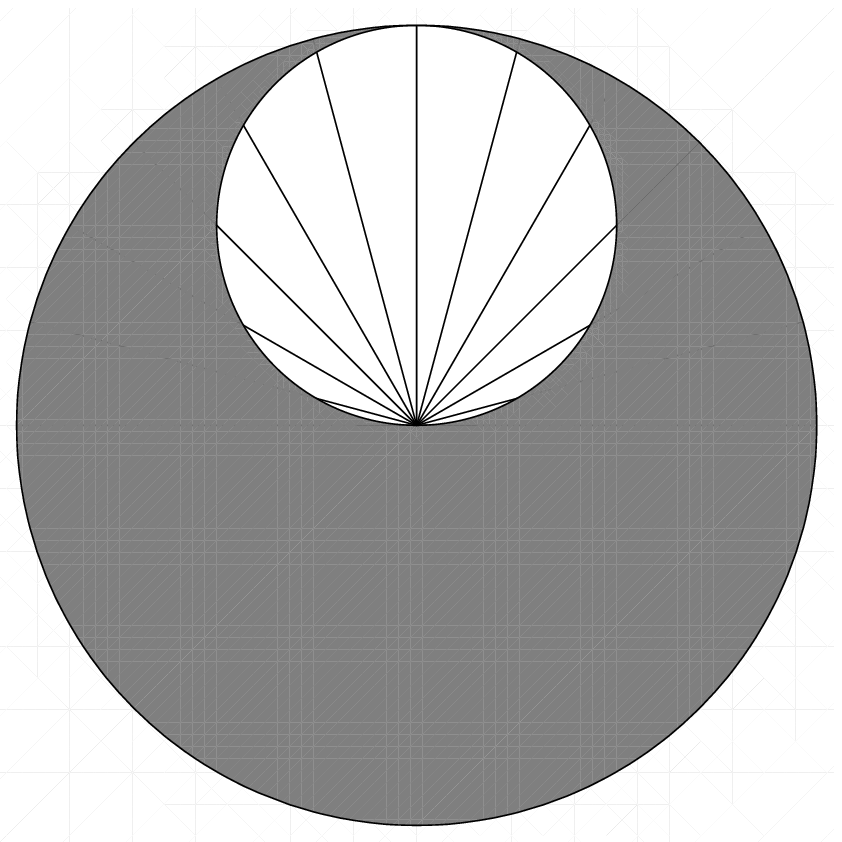}} \caption{$\sigma_c = 1, \sigma_w = 0$ \label{fig1}}
\end{minipage}
\hfill 
\begin{minipage}[t]{0.46\linewidth}
\centering  {\includegraphics[height=2.5in,
width=2.5in]{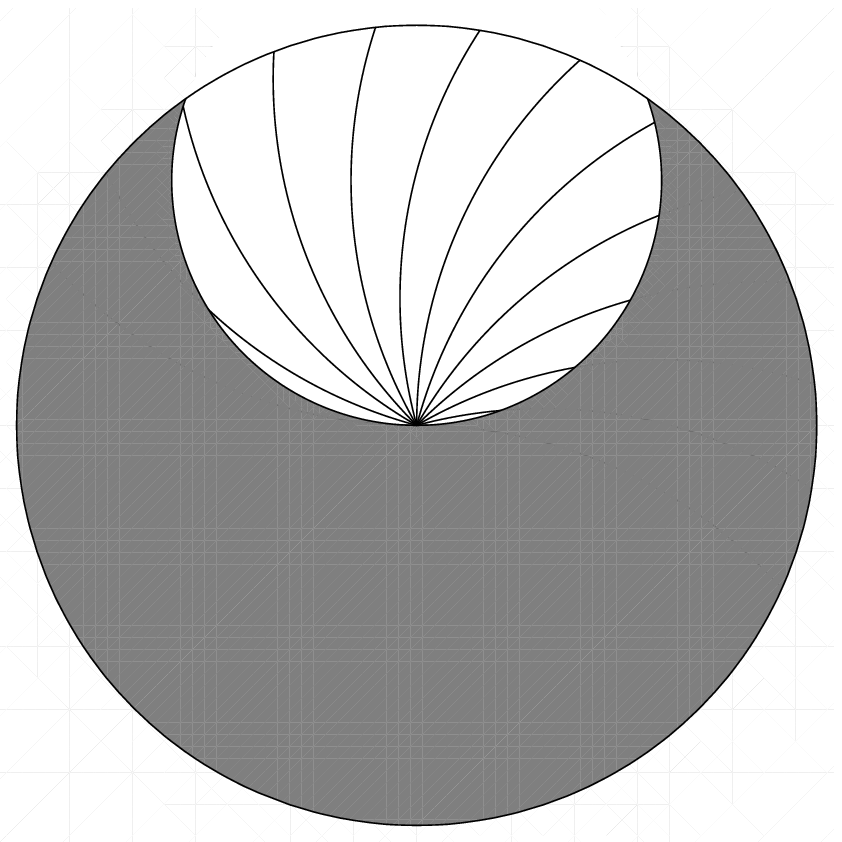}} \caption{$\sigma_c = 1, \sigma_w = .5$ \label{fig2}}
\end{minipage}
\end{figure}

\begin{figure}[!h]
\begin{minipage}[t]{0.46\linewidth} 
\centering {\includegraphics[height=2.5in,
width=2.5in]{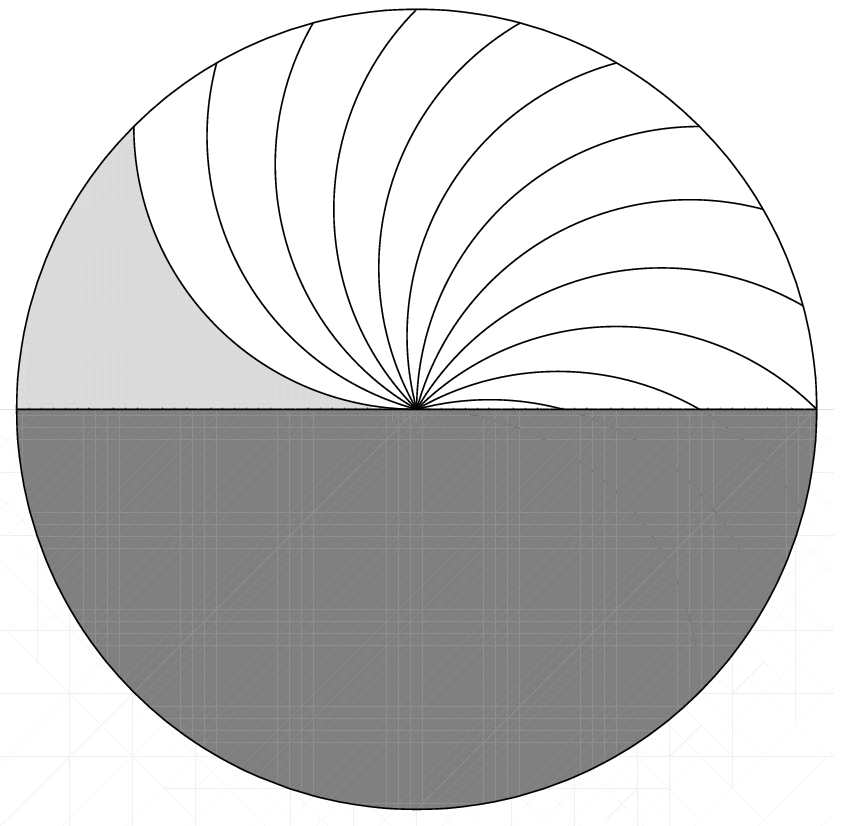}}\caption{$\sigma_c = 0, \sigma_w = .5$\label{fig3}}
\end{minipage}
\hfill 
\begin{minipage}[t]{0.46\linewidth}
\centering  {\includegraphics[height=2.5in,
width=2.5in]{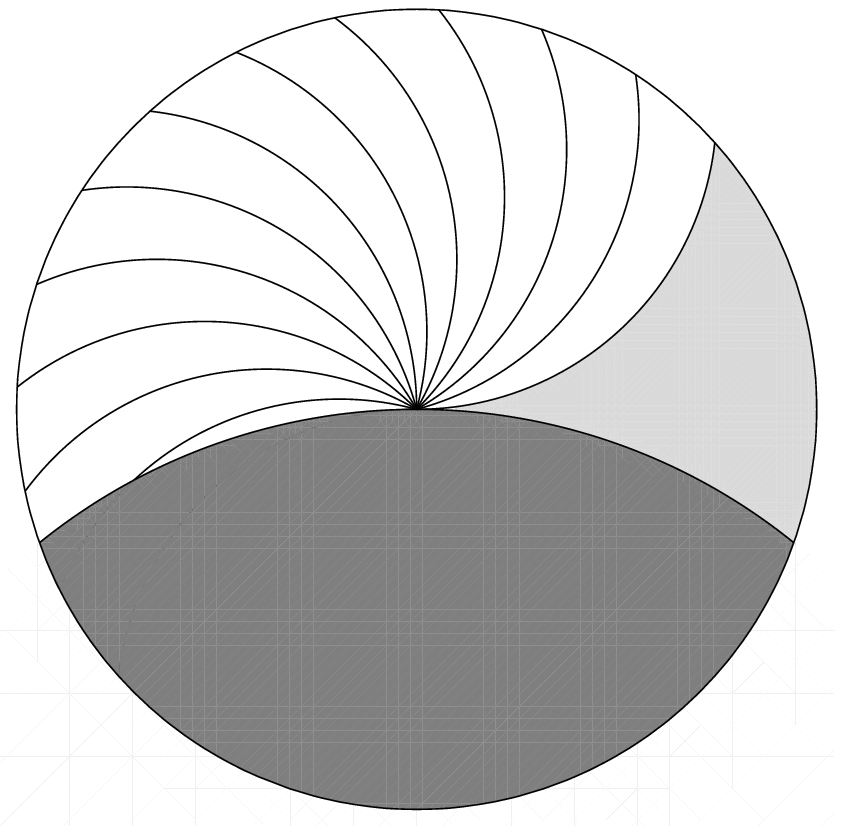}} \caption{$\sigma_c =- 1, \sigma_w = -2$\label{fig4}}
\end{minipage}
\end{figure}

We will assume\footnote{this is easily arranged by making use of
  symmetries of the Poincar\'e disk.} that the source of the noise is
located at the centre of the disk. We then have the following building
blocks for our geometric construction
\begin{itemize}
\item \textbf{Ground level} corresponds to an arc of a circle of radius 
\begin{equation}
\frac{1}{2}\sqrt{1+2 \frac{\sigma_w^2}{\sigma_c^2}}
\end{equation}
through the origin. If $\sigma_c>0$, so that the speed of sound increases with
height, the atmosphere corresponds to the interior of this
circle otherwise it is the exterior. By a rotation we may arrange
that at $z=0$ the ground level is tangent to $\textrm{Im}{z}=0$, with the
atmosphere above the ground. 
\item \textbf{Sound Rays} are arcs of circles of radius
\begin{equation}
\frac{1}{2}\sqrt{2+ \frac{\sigma_c^2}{\sigma_w^2}}
\end{equation}
passing through the origin. If $\sigma_w>0$, so that the wind blows
from left to right above ground-level, the arcs should curve to the
right, else they should curve to the left.
\end{itemize}
Figures  \ref{fig1} to  \ref{fig4} show sample constructions for various values of
$\sigma_c$ and $\sigma_w$. The ground is shaded, and rays are shown
emanating from the origin at equally spaced angles of $\pi/12$. Since
the Poincar\'e model is conformally flat, a source which radiates
uniformly in all directions will radiate the same amount of energy
between any two rays, $1/12$ of the total energy radiated. We assume
for simplicity that there is no significant transfer of energy between
the bulk fluid motion and the sound waves.

Figure \ref{fig1} shows the case of vanishing wind, together with a sound speed
which increases with height, such as may be caused by a `temperature
inversion'. In this case the rays are straight lines and all return to the
ground at some point. Clearly one half of the power radiated returns
to each side of the source.

Figure \ref{fig2} shows the case where sound speed increases with height and
there is also a wind shear, with the wind blowing from left to
right. In this case, not all the radiated energy returns to earth. By
counting the rays, we readily estimate that roughly $5/12$ of the
power is lost to the atmosphere, $5/12$ is received downwind (i.e.\ to
the right) and
$2/12$ upwind (left) of the source.

Figure  \ref{fig3} shows the case where the sound speed does not vary
significantly with height, but there is a wind blowing left to
right. In this case, in the ray theory approximation\footnote{We would expect that for a full solution of the wave
equation the sound field will not be zero but will in fact vanish
exponentially in any `silent' zone of the ray theory approximation.}, no sound is received to the left of the source, approximately $1/4$
is received to the right and the rest is radiated upwards to the
atmosphere. 

Figure  \ref{fig4} shows the case where there is a decrease in sound speed with
height, together with a strong wind shear, with the wind blowing from right to left. In this
case there are competing effects between the changing sound speed,
which would tend to refract waves upwards and the wind shear which
would tend to bend rays downwards. We see that to the right of the source there is no
sound received, while roughly $1/8$ of the power is received
to the left. In both Figure  \ref{fig3} and Figure  \ref{fig4} there is a `quiet zone', shown
shaded in light grey, where no sound reaches the observer.

\begin{figure}[!h]
\begin{minipage}[t]{0.46\linewidth} 
\centering {\includegraphics[height=1.25in,
width=2.5in]{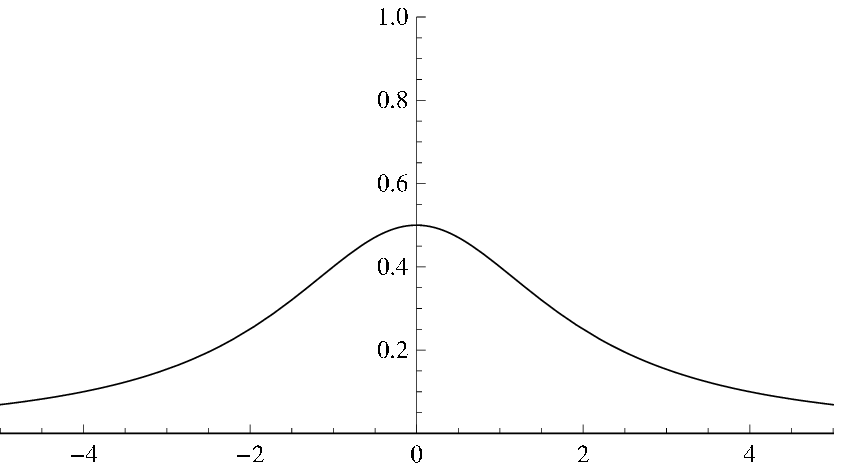}} \caption{$\sigma_c = 1, \sigma_w = 0$ \label{fig5}}
\end{minipage}
\hfill 
\begin{minipage}[t]{0.46\linewidth}
\centering  {\includegraphics[height=1.25in,
width=2.5in]{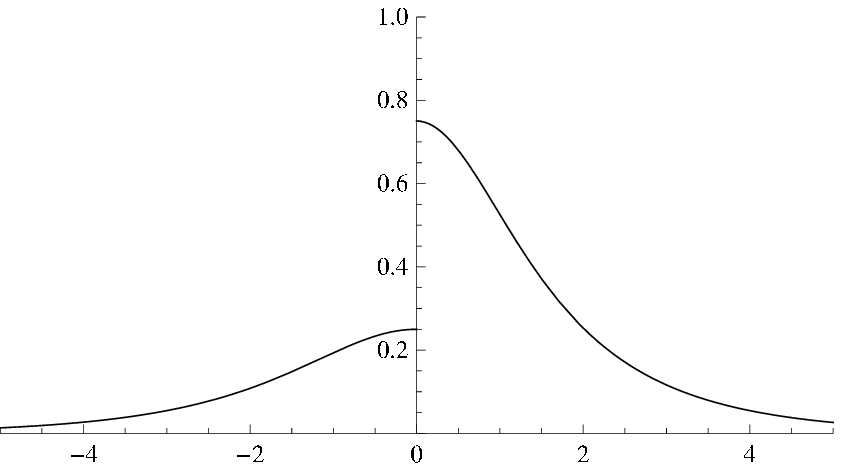}} \caption{$\sigma_c = 1, \sigma_w = .5$ \label{fig6}}
\end{minipage}
\end{figure}

\begin{figure}[!h]
\begin{minipage}[t]{0.46\linewidth} 
\centering {\includegraphics[height=1.25in,
width=2.5in]{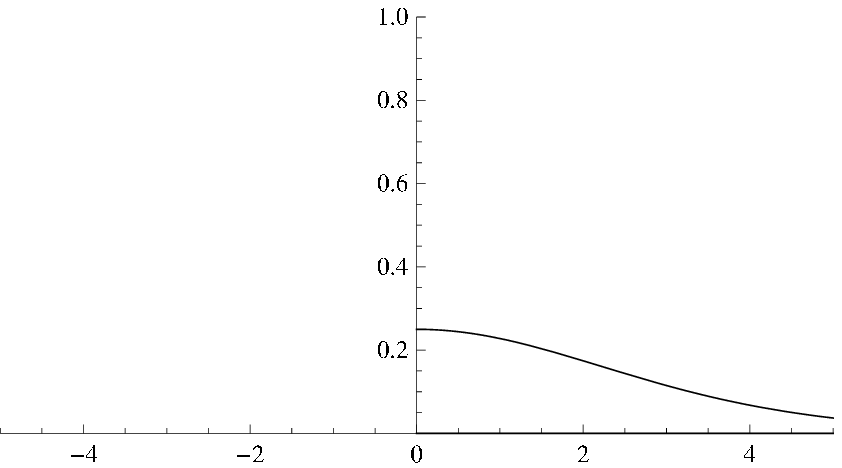}}\caption{$\sigma_c = 0, \sigma_w = .5$\label{fig7}}
\end{minipage}
\hfill 
\begin{minipage}[t]{0.46\linewidth}
\centering  {\includegraphics[height=1.25in,
width=2.5in]{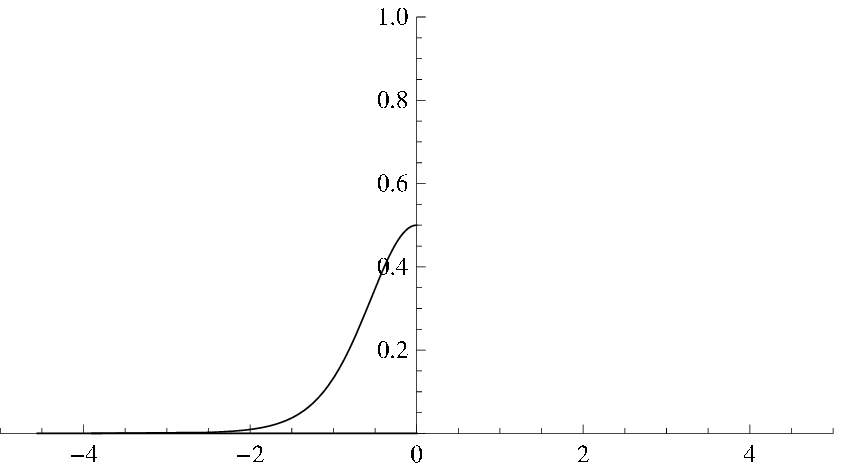}} \caption{$\sigma_c =- 1, \sigma_w = -2$\label{fig8}}
\end{minipage}
\end{figure}

We see then how a simple `straightedge and compass' construction allows us
to make quantitative predictions about the ratio of power transmission
upwind and downwind in a shearing wind. In fact, with a little more
geometry it is possible to calculate, in our
approximation, the intensity of sound received
as a function of distance from the source. Suppose a sound ray is emitted from the source at an
angle $\theta$ to the ground. It is a matter of simple circle geometry
to calculate the point at which this ray again intersects the
ground. Using the hyperbolic metric we can find the proper distance
along the ground to this point, and we find that it is given by
\begin{equation}
x = c_0 \frac{\sqrt{2}}{\sigma_w} \tanh^{-1}\left(\frac{\sqrt{2}
  \sigma_w \sin \theta}{\sigma_w + \sigma_c \cos \theta }\right),
\end{equation}
where $c_0=c(0)$. The received intensity is then simply proportional to
$\left(\frac{dx}{d\theta}\right)^{-1}$. We plot $x$ against
$\left(\frac{dx}{d\theta}\right)^{-1}$ for the same values of
$\sigma_c, \sigma_w$ as above in Figures \ref{fig5} to \ref{fig8}.

In these figures, we have set $c_0=1$. The vertical
scale is proportional to the intensity received. The plots agree with
the rough ratios we found above for the energy upwind and downwind of
the source.

The approximations we
have made are valid provided that $\sigma_c x / c_0$ and $\sigma_w x /
c_0$ are small, so we can trust the plots in a neighbourhood of the
origin. Qualitatively, we do not expect significant deviation from
these plots, even outside the regime where the approximations are
valid.

We note that Randers metrics and Zermelo problems are in direct, one-to-one
correspondence. We might therefore choose to take the view that
the preceding calculations are exact, but for an atmosphere with a
slightly different sound and wind speed profile. These profiles would
match the assumed shape above in a region near the ground.

\section{Ray Plots \label{rays}}

\begin{figure}[!t]
\centering {\includegraphics[height=1.5in,
width=4.5in]{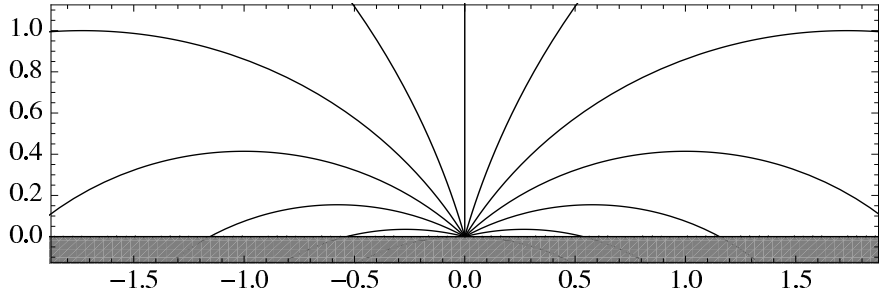}}\caption{$\sigma_c = 1, \sigma_w = 0$\label{fig9}}
\end{figure}

\begin{figure}[!t]
\centering {\includegraphics[height=1.5in,
width=4.5in]{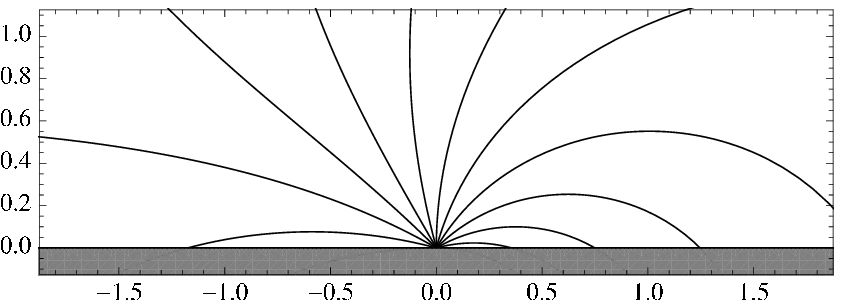}}\caption{$\sigma_c = 1, \sigma_w = 0.5$\label{fig10}}
\end{figure}

\begin{figure}[!t]
\centering {\includegraphics[height=1.5in,
width=4.5in]{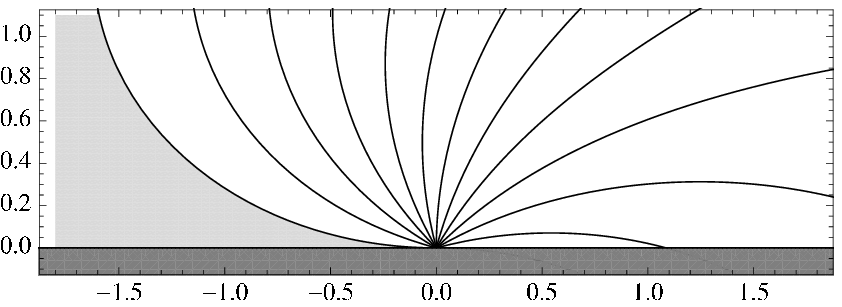}}\caption{$\sigma_c = 0, \sigma_w = 0.5$\label{fig11}}
\end{figure}

\begin{figure}[!t]
\centering {\includegraphics[height=1.5in,
width=4.5in]{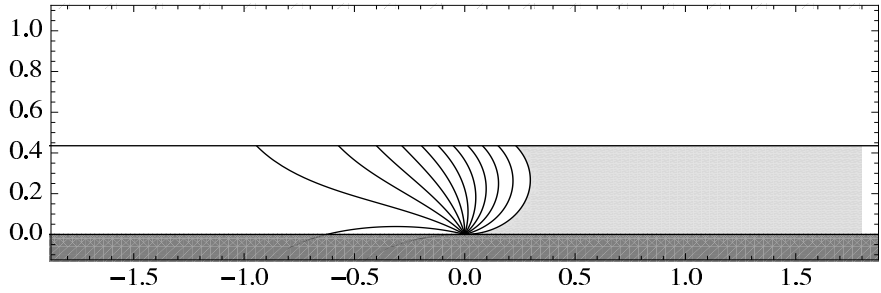}}\caption{$\sigma_c = -1, \sigma_w = -2$\label{fig12}}
\end{figure}

In order to plot the paths of rays in the physical $x, z$ coordinates,
we need to explicitly construct the map to a region of the hyperbolic
disk, of which we have made implicit use in the preceding section. We do this in the appendix below. In this section we use the mapping to pull back the ray paths from the hyperbolic disk to the original coordinates so as to exhibit the physical paths of the rays. Figures \ref{fig9} to \ref{fig12} show these paths for the same parameters as the previous section.

We once again take $c_0=1$. We note that the ray paths are broadly as
 one would expect from heuristic considerations \cite{Stokes, Reynolds}.  In Figure \ref{fig12} we explicitly see that the rays terminate at $z \approx .42$. The horizontal line here is the pull back of the conformal boundary of $\mathbb{H}^2$, and the rays will take an infinitely long time to reach this height. Figures \ref{fig10} and \ref{fig11} would also exhibit such a phenomenon if the $z$-axis were extended.
 In practice of course, the approximations we have made are only valid in a strip about $z=0$ and would break down before the rays reached this height.
 
 One may verify roughly that these figures are consistent with the intensity plots given above. The energy absorbed by the ground between any adjacent rays is $\pi/12$, which should be the area under the intensity plot between the same values of $x$.

\section{The sound field of a moving source \label{mach}}

\begin{figure}[!ht]
\centering {\includegraphics[height=1.5in,
width=4.5in]{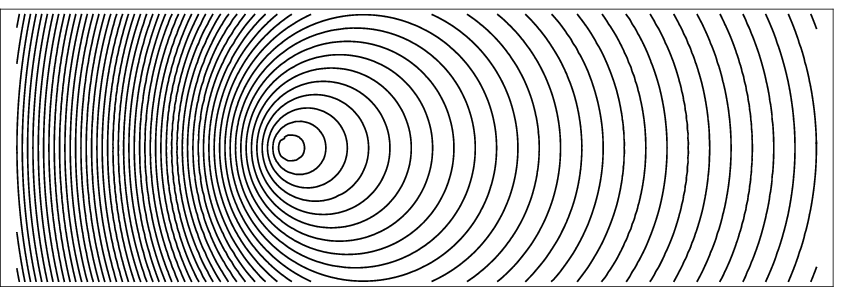}
\includegraphics[height=1.5in,
width=4.5in]{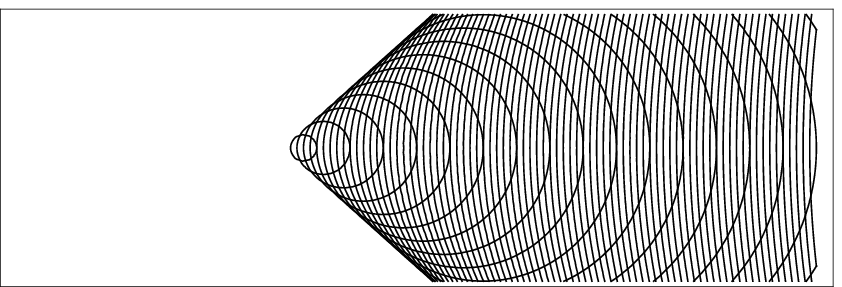}
}\caption{Example sound fields for a subsonic (t) and supersonic (b) body in a
  homogeneous atmosphere with no wind\label{fig13}}
\end{figure}

We now change perspective somewhat and leave behind the problem
of propagation of traffic noise to instead consider a moving
body which radiates sound, for example an aircraft. In the classical
case of a spatially
homogeneous $2$-dimensional atmosphere with no wind, a subsonic aircraft moving along the
$z$-axis at constant speed has a
soundfield that fills the whole of $\mathbb{R}^2$. For an aircraft
moving at a supersonic speed, the disturbances are confined inside the
`Mach cone' and there is a zone of silence in front of the body. There
is a `Sonic Boom' at the boundary of this cone. These features are
shown in Figure \ref{fig13}. It is known that the introduction
of a spatially varying sound speed gives rise to 
interesting, qualitatively different phenomena in the sound fields of
fast moving objects \cite{Kaouri}. We shall examine the further
effects to be observed in the presence of a wind.

In order to
visualise the sound field, it is convenient to imagine that the object
emits pulses of sound periodically, and to plot the loci of all these
disturbances at some fixed time. We work in the approximation outlined
in Section \ref{finssec}, where the wavelength of sound is assumed to be much shorter
than other lengthscales in the problem and so the disturbances
propagate along geodesics of the Randers metric $F$ of (\ref{fins}, \ref{ztor}). The
locus of a disturbance emitted a time $t$ in the past is therefore a
geodesic circle of radius $t$ with respect to the Randers metric $F$,
centred on the point at which the disturbance originated.

We shall consider a body moving in the stratified atmosphere of
Sections \ref{trafnoise}, \ref{rays} at a constant speed parallel to
the $x$-axis. After a Galileian boost, we may assume that the wind
speed is zero at the height of the body, so that it moves along $z=0$
without loss of generality. In order to visualize the sound field as
described above, we need to find the geodesic circles of the Randers
metric defined by
\begin{equation}
a = \frac{dx^2}{c^2(z)\left(1-\frac{w^2(z)}{c^2(z)}
  \right)^2}+\frac{dz^2}{c^2(z)\left(1-\frac{w^2(z)}{c^2(z)} \right)},
  \qquad b=-\frac{w(z)}{c^2(z)-w^2(z)}dx, \label{stratrand}
\end{equation}
centred on points of $z=0$. Since $\partial/\partial x$ is a
Killing vector of the Randers metric, we may without loss of
generality consider only those geodesic circles centred on the
origin. 

We once again assume that $w(z)/c(z) \ll 1$ and $c''(z) c(z)/c'(z)^2\ll
1$, so that the geodesics of the Randers metric are well approximated
by the rays found in previous sections. We note that the time taken to
traverse a curve $\gamma(\lambda)$ has two contributions:
\begin{equation}
t = \int d\lambda (\sqrt{a_{ij} \dot{\gamma}^i\dot{\gamma}^j}+b_i \dot{\gamma}^i)
\end{equation}
but that (\ref{stratrand}) shows the $b_i \dot{\gamma}^i$ contribution
to be smaller than the first contribution by a factor of $w/c$. We may
thus approximate the length of a curve with respect to the Randers
structure $a+b$ by the Riemannian length of the metric $a$\footnote{It might
seem that we have removed any possible effect the wind might have by
such an assumption. It should be noted however that the geodesics
themselves depend on the wind. In this approximation, we keep the
\emph{refractive} effect of wind shear, while ignoring the
\emph{transport} effect of the wind, which will be negligible for
reasonable wind speeds.}. We shall
find the geodesic circles of radius $t$ by moving a
distance $t$, with respect to $a$, along the geodesics calculated in
previous sections. We make use of the approximations above to map the
problem to the hyperbolic disk, where the manifest rotational
symmetries of the problem show that the geodesic circles we seek are
in fact Euclidean circles centred on the origin. The relation between
the Euclidean radius and the Randers radius is found by integrating
$a$ along an arc of a circle. Finally the transformations given
explicitly in the Appendix are used to map the geodesic circle back to
physical space. Combining this construction with the $x$-translation
invariance of the physical space, we may readily plot the
sound field for various values of the parameters $\sigma_c$,
$\sigma_w$ and for differing speeds of the radiating source, expressed
as a Mach number as $v = M c_0$, with $c_0$ the speed of sound at $z=0$.

\begin{figure}[!ht]
\centering {\includegraphics[height=1.5in,
width=4.5in]{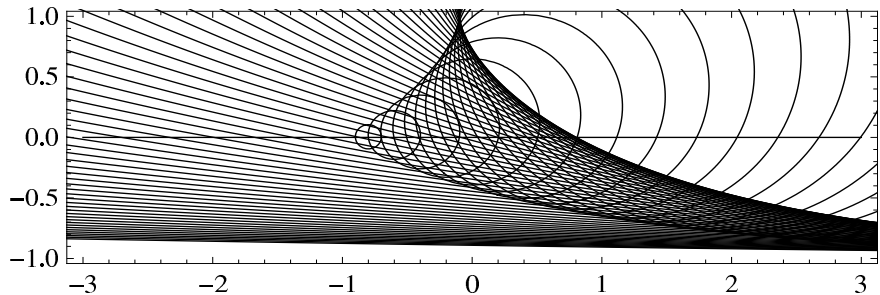}
}\caption{Sound field for a body with $M=2$ in an atmosphere with $\sigma_c=1$, $\sigma_w=0$\label{fig14}}
\end{figure}

\begin{figure}[!ht]
\centering {\includegraphics[height=1.5in,
width=4.5in]{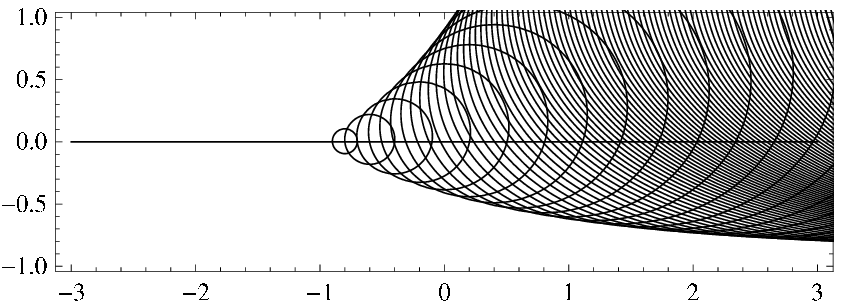}
\includegraphics[height=1.5in,
width=4.5in]{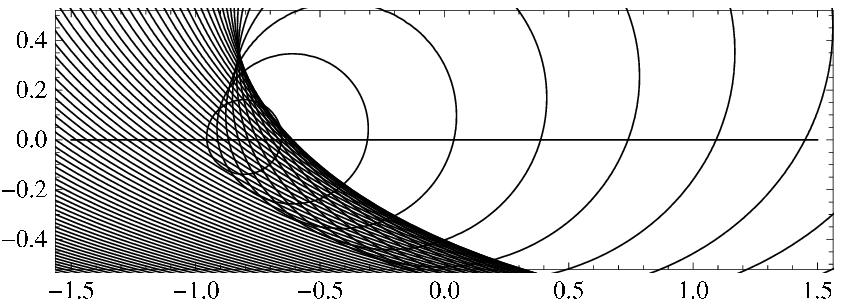}
}\caption{Sound field for a body with $M=2$ (t) and $M=1.3$ (b) in an
  atmosphere with $\sigma_c=1$, $\sigma_w=0.5$. Note the altered range
  of the lower figure.\label{fig15}}
\end{figure}

\begin{figure}[!ht]
\centering {\includegraphics[height=1.5in,
width=4.5in]{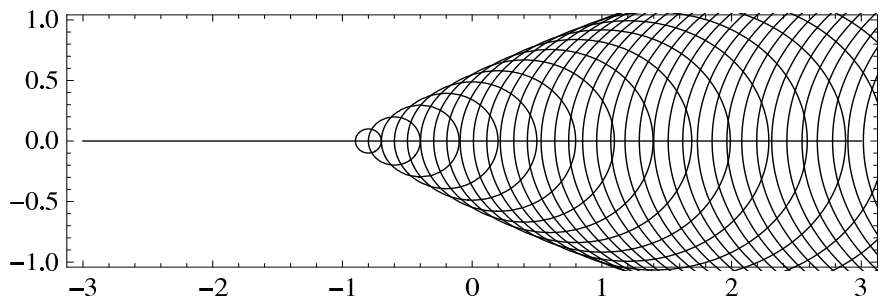}
}\caption{Sound field for a body with $M=2$ in an atmosphere with $\sigma_c=0$, $\sigma_w=0.5$\label{fig16}}
\end{figure}

Figures \ref{fig14} to \ref{fig16} show some sample sound fields. We
see that Figure \ref{fig14} is qualitatively very similar to that of
\cite{Kaouri} which considers a moving source in a stratified
atmosphere whose speed of sound varies as $c(z) = (1-z)^{-1/2}$. This
particular sound speed profile is chosen do that the rays are
parabolae with vertical axes. For
such an atmosphere, there is no `zone of silence', essentially because
rays which travel into the upper half plane and are refracted back
downwards can `overtake' the moving source. This is because the source
is moving at a speed which is supersonic only with respect to the
local sound speed, whereas in this atmosphere the speed of sound
increases with height so that a ray travelling at a sufficient height
will travel faster than the source. There is however a `Mach
envelope', so an observer may experience a sonic boom. The envelope
has a cusp at the height at which the local sound speed equals the
speed of the source.

Figure \ref{fig15} shows sound fields for a supersonic object moving
in a stratified atmosphere with both a varying speed of sound and a
wind. We find that there are two regimes. For a source moving much
faster than the speed of sound, the field is similar to the Mach cone
of a supersonic body in a homogeneous wind-free atmosphere, albeit
distorted by the inhomogeneity. For a source which moves more slowly,
but still above the local speed of sound, the picture is more similar
to that of Figure \ref{fig14}. The reason for these different regimes
is that the wind shear and varying speed of sound produce competing
effects. The varying speed of sound tends to bend rays above the axis
downwards, while the wind shear tends to bend rays travelling from
right to left upwards. For a source travelling just over the local
speed of sound, rays which travel into the upper half-plane and are
refracted down may overtake the object as in Figure \ref{fig14}. For a
source travelling considerably faster than the local speed of sound
this cannot happen as any ray which moves far enough into the upper
half-plane so that it moves faster than the source is refracted by the
wind shear and never returns to $z=0$.

Figure \ref{fig16} shows the case of an atmosphere whose sound speed
is constant, but which has a varying wind. In this case, the sound
field is similar to the Mach cone of the homogeneous wind-free case,
but that the cone bends inwards slightly away from the axis due to the
refraction caused by the wind shear.

\section{$SL(2, \mathbb{R})$, squashed AdS$_3$ spacetimes and integrability}

In \cite{Gibbons:2008zi} we explored a third vertex to the correspondence
between the Zermelo problem and the problem of finding geodesics of a
Randers metric. Both problems, and the correspondence between them,
fit naturally into the problem of finding the null geodesics of a
conformally stationary Lorentzian spacetime. Firstly we note that any
equivalence class of metrics $[g]$ which are conformally stationary
locally have a representative of the form
\begin{equation}
g_R = -(dt-b_i(x)dx^i)^2+a_{ij}(x) dx^i dx^j
\end{equation}
with $[g_R]=[g]$, which we call the Randers form of the metric. If we choose to parameterize the null geodesics of
$[g]$ by $t$, then they are in fact geodesics of the Randers metric
\begin{equation}
F(x,y)  = \sqrt{a_{ij}(x)y^i y^j}+b_iy^i \label{newrand}
\end{equation}
parameterized by unit Finslerian length. This choice of representative
naturally picks out a Riemannian metric $a$ together with a one-form
$b$. 

Another way to write a manifestly stationary spacetime, often referred
to as the Painlev\'e-Gullstrand form, is given by
\begin{equation}
g_Z = -dt^2 + h_{ij}(x)\left(dx^i + W^i(x) dt \right)\left(dx^j +
W^j(x) dt \right). \label{zermform}
\end{equation}
We show in \cite{Gibbons:2008zi} that $[g_Z]=[g_R]$ if and only if the
Randers metric (\ref{newrand}) solves the Zermelo problem with data
$(h_{ij}, W^i)$. The condition that the Randers metric be strongly
convex is also necessary to ensure that the conformal factor is
nowhere singular. We thus see that the Randers and Zermelo pictures
arise as different ways to write a conformally stationary class of
conformal metrics.

We now have a prescription to lift the Finslerian geometry of the
Randers structure to the more familiar Lorentzian geometry. The
problem we have explored in depth above is that of a constant magnetic
field $\beta$ on the hyperbolic plane with curvature $-\kappa^2$. Following \cite{Gibbons:2008zi}, we
note that the lifted metric in the Randers form may be written
\begin{equation}
g_R = \frac{-\beta^2
\left(\rho^1\right)^2+\left(\rho^2\right)^2+\left(\rho^3\right)^2}{\kappa^2} \label{rinvmet}
\end{equation}
where the $\rho_i$ are right invariant one-forms on
$SL(2,\mathbb{R})$:
\begin{eqnarray}
\rho^1 &=& \frac{\kappa dt}{\beta}+\frac{1+r^2}{1-r^2} d\phi \nonumber\\
\rho^2 &=& \frac{2}{1-r^2}\left(r \cos (\kappa t/\beta) \, d\phi -
\sin(\kappa t/\beta) \, dr \right)\nonumber\\
\rho^3 &=& \frac{2}{1-r^2}\left(r \sin (\kappa t/\beta) \, d\phi +
\cos(\kappa t/\beta) \, dr \right). \label{rightinv}
\end{eqnarray}
These arise by parameterizing $U \in SL(2, \mathbb{R})$ as
\begin{equation}
U=\left( \begin{array}{cc}
W+X & V-Y  \\
-V-Y & W-X   \end{array} \right), \qquad \begin{array}{c}
X+i Y = \frac{i
  r}{\sqrt{1-r^2}} e^{i (\kappa t-\beta \phi)/(2 \beta)}, \\  W+i V =
\frac{1}{\sqrt{1-r^2}} e^{i (\kappa t+\beta \phi)/(2 \beta)}. \end{array}
\end{equation}
The standard procedure of writing the Maurer-Cartan form
\begin{equation}
dU U^{-1} = \rho^i \tau_i,
\end{equation}
with $\tau_i$ a basis for $sl(2, \mathbb{R})$ given in terms of the
standard Pauli matrices as $\{\frac{i}{2} \sigma_2,
\frac{1}{2}\sigma_3, -\frac{1}{2} \sigma_1 \}$ yields
(\ref{rightinv}). 

Thus the sound rays we found are projections of null
geodesics of a right-invariant metric on $SL(2, \mathbb{R})$. The ease
with which we were able to construct the rays is then seen to be a
consequence of symmetry. The metric (\ref{rinvmet}) certainly
admits all the left-invariant vector fields of $SL(2, \mathbb{R})$ as
Killing fields. These are readily constructed as the dual basis to the left-invariant
one-forms which satisfy $U^{-1} dU = \lambda^i \tau_i$ and are given
by:
\begin{eqnarray}
L_1 &=& \pd{}{\phi} \nonumber \\
L_2 &=& \frac{1+r^2}{2r}\cos\phi\pd{}{\phi} + \frac{1-r^2}{2r} \left(r \sin
\phi \pd{}{r} - \beta \cos\phi \pd{}{t} \right)\nonumber \\
L_3 &=& -\frac{1+r^2}{2r}\sin\phi\pd{}{\phi} + \frac{1-r^2}{2r} \left(r \cos
\phi \pd{}{r} + \beta \sin\phi \pd{}{t} \right).
\end{eqnarray}
In addition, since the coefficients of $\rho^2, \rho^3$ in the metric
are the same, there is a further Killing field
\begin{equation}
R_1 = \beta\pd{}{t},
\end{equation}
so that the full symmetry group of the metric is $(SL(2, \mathbb{R}) \times \mathbb{R} )/ \mathbb{Z}_2$. The $\mathbb{Z}_2$ quotient cannot be seen at the level of the Lie algebra, but one sees upon exponentiating the algebra that it is in fact a double covering of the symmetry group. Using the isomorphism
\begin{equation}
SO(2,1) \cong SL(2, \mathbb{R}) / \mathbb{Z}_2
\end{equation}
we can also identify the symmetry group as $SO(2,1) \times \mathbb{R}$ with the $SO(2,1)$ group representing the symmetries of the hyperbolic plane and the factor $\mathbb{R}$ the time translation symmetry. There are many similarities between the situation we are dealing with and that of right-invariant metrics on $S^3 \cong SU(2)$, which should come as no surprise since the groups are closely related by analytic continuation. The metrics we have exhibited are a natural analogue of the biaxial Berger metrics on $S^3$, with symmetry group $(SU(2) \times U(1)) / \mathbb{Z}_2$.

The metrics (\ref{rinvmet}) are often referred to as describing a
squashed anti-de Sitter geometry \cite{Rooman:1998xf}. For $\beta^2 > \kappa^2$ we have a family of
metrics similar to (a factor of) the G\"odel universe, which include
the G\"odel universe\footnote{strictly speaking the G\"odel
  universe is the product of this space with a flat direction} as the special case $\beta^2 =2 \kappa^2$ and
which all admit closed timelike curves (CTCs). For $\beta^2<\kappa^2$ the
spacetimes do not admit CTCs.

After approximation, we found above that the problem of finding
sound rays for a stratified atmosphere with a wind shear is equivalent
to that of finding the motion of a charge in a constant magnetic field
of $\beta=\sigma_w$ in a hyperbolic space with $\kappa^2 =
\sigma_c^2+2\sigma_w^2$, with $\sigma_c = c'(0),\, \sigma_w =
w'(0)$. Clearly we will always be in the regime where $\beta<\kappa$
so do not expect any CTCs in the lifted spacetime. This is related to
the fact that any two points in the space may be joined by
a Randers geodesic and these properties are explored in more detail in
\cite{Gibbons:2008zi}.

\section{Conclusion}

We have examined the old problem of ray tracing for sound waves in a
wind. We find that in a certain physically plausible approximation the
ray paths are mapped to the trajectories of a charged
particle moving in a uniform magnetic field on the hyperbolic
plane. We have exploited this fact to produce plots of the intensity
of sound received from a point source at ground level in the presence
of a wind and varying speed of sound. We have also constructed the
approximate Mach envelopes for a moving body in the presence of a
wind. We have discussed the symmetries of this model and how they
relate to the problem of finding null geodesics of a family of right invariant
metrics on $SL(2, \mathbb{R})$ which are often referred to as squashed
AdS metrics.

\section*{Acknowledgements}

GWG would like to thank Hugh Hunt for an enlightening discussion. We
would also like to thank Michael Berry and John Ockendon for interesting
comments on the manuscript.

\appendix

\section*{Appendix - Mapping to $\mathbb{H}^2$.}

We start with the metric $a$ from above:
\begin{equation}
a = \frac{dx^2}{c^2(z)\left(1-\frac{w^2(z)}{c^2(z)}
  \right)^2}+\frac{dz^2}{c^2(z)\left(1-\frac{w^2(z)}{c^2(z)} \right)},
\end{equation}
and make a near-identity change\footnote{in practice we may ignore this change of variables as $y=z$ to $O(y^3)$, which is the order of the approximations we make} of the vertical coordinate
\begin{equation}
dy =\sqrt{1-\frac{w^2(z)}{c^2(z)}} dz
\end{equation}
so that the metric may be written in the form
\begin{equation}
a = \frac{dx^2+dy^2}{c^2(y)\left(1-\frac{w^2(y)}{c^2(y)}\right)^2} = \frac{dx^2+dy^2}{f^2(y)}.
\end{equation}
At this point we will explicitly approximate $f(y)$ as
\begin{equation}
f(y) = \alpha \cos(\beta y- \gamma)
\end{equation}
which for
\begin{equation}
\alpha = c_0\left(1+\frac{\sigma_c^2}{2\sigma_w^2} \right)^{\frac{1}{2}}, \qquad \beta = \frac{\sigma_w}{c_0} \sqrt{2}, \qquad \gamma = \sin^{-1}\left(\frac{\sigma_c}{\alpha \beta} \right)
\end{equation}
is accurate to $O(y^3)$. After a shift in the $y$ coordinate defined by
\begin{equation}
\tilde{y} = y-\frac{\gamma}{\beta}
\end{equation}
we have
\begin{equation}
a = \frac{dx^2+d\tilde{y}^2}{\alpha^2 \cos^2 \beta y} = \frac{d\zeta d\bar{\zeta}}{\alpha^2 \cos^2 \left[ \beta(\zeta-\bar{\zeta})/2 i\right]}
\end{equation}
where we define $\zeta = x + i \tilde{y}$. This is a (perhaps unusual) metric for the hyperbolic plane of curvature $-\alpha^2 \beta^2$, where the strip $-\pi/(2\beta) < \tilde{y} < \pi/(2\beta)$ is a complete copy of the plane. We map this strip onto the standard unit disk with the conformal mapping
\begin{equation}
\omega = \tanh\frac{\beta \zeta}{2}
\end{equation}
so that the metric $a$ becomes
\begin{equation}
a = \frac{4}{\alpha^2 \beta^2} \frac{d\omega d\bar{\omega}}{\left( 1-\abs{\omega}^2\right)^2}
\end{equation}
The ground is mapped to a curve of constant geodesic curvature passing through $\omega=\pm 1$. A final M\"obius transformation leaves the metric unchanged, but shifts the ground so that it passes through the origin
\begin{equation}
\omega' = \frac{p \omega + i}{-i \omega + p}, \qquad \textrm{where} \quad p=\sqrt{2}\frac{\sigma_w}{\sigma_c}+\sqrt{1+2 \frac{\sigma_w^2}{\sigma_c^2}}.
\end{equation}


\begin{thebibliography}{1}

\bibitem{Stokes}
  G.~G.~Stokes, ``On the Effect of Wind on the Intensity of Sound,''
  Report of the British Association, Dublin, 1857, p22. [Also
  in Mathematical and Physical Papers by George Gabriel Stokes,
  Vol. IV, CUP, Cambridge (1880) p110]

\bibitem{Reynolds}
  O.~Reynolds, ``On the Refraction of Sound by the Atmosphere,''
  Proc. Roy. Soc. {\bf 22} (1874) p531.

\bibitem{Rayleigh}
J.~W.~S.~Rayleigh, ``The Theory of Sound'', Macmillan (1986)  \S289, Vol 2.

\bibitem{Milne}
E.~A.~Milne, ``Sound Waves in the Atmosphere,'' Phil. Mag. Series 6 (1921),
{\bf 42}:247 96

\bibitem{Kornhauser}
E.~T.~Kornhauser, ``Ray theory for Moving Fluids, '' Journal of the
Acoustical Society of  America {\bf 25} (1953) 945

\bibitem{Jones}
  M.~R.~Jones, E.~S.~Gu, and A.~J.~Bedard, Jr., ``Infrasonic
  Atmospheric Propagation Studies Using a 3-D Ray Trace Model'',
  Proceedings, 22nd Conference on Severe Local Storms, 2004.

\bibitem{Gibbons:2008zi}
  G.~W.~Gibbons, C.~A.~R.~Herdeiro, C.~M.~Warnick and M.~C.~Werner,
  ``Stationary Metrics and Optical Zermelo-Randers-Finsler Geometry,''
  Phys.\ Rev.\  D {\bf 79} (2009) 044022
  [arXiv:0811.2877 [gr-qc]].

\bibitem{BaoRoblesShen}
D.~Bao, C.~Robles and Z.~Shen,
``Zermelo navigation on Riemannian manifolds,''
J.\ Diff.\ Geom.\ {\bf 66} (2004) 377-435.

\bibitem{MeyerSchroeter} 
R.~Meyer and G.~Schroeter, 
``The application of differential geometry to ray acoustics in inhomogeneous moving media,''
Acustica {\bf 47} (1981) 105.

\bibitem{Berry}
M.~V.~Berry, R.~G.~Chambers, M.~D.~Large, C.~Upstill and J.~C.~Walmsley
``Wavefront dislocations in the Aharonov-Bohm effect and its water wave analogue,''
Eur.J.Phys {\bf 1} (1980), 154-162.

\bibitem{White} 
R.~White, 
``Acoustic ray tracing in moving inhomogeneous fluids,''
Journal of the Optical Society of  America {\bf 53} (1973) 1700.

\bibitem{Kaouri}
K.~Kaouri, D.J.~Allwright, C.J.~Chapman and J.R.~Ockendon,
``Singularities of wavefields and sonic boom,''
Wave Motion {\bf 45} (2008) 217-237.

\bibitem{Rooman:1998xf}
  M.~Rooman and P.~Spindel,
  ``Goedel metric as a squashed anti-de Sitter geometry,''
  Class.\ Quant.\ Grav.\  {\bf 15} (1998) 3241
  [arXiv:gr-qc/9804027].


\end{thebibliography}
\end{document}